# Cavity QED with Diamond Nanocrystals and Silica Microspheres

Young-Shin Park, Andrew K. Cook, and Hailin Wang[*]

Department of Physics, University of Oregon, Eugene, Oregon 97403, USA

[*]Corresponding author. Tel: 1-541-346-4758. Fax: 1-541-346-4315. E-mail: hailin@uoregon.edu

**Abstract**   Normal mode splitting is observed in a cavity QED system, in which nitrogen vacancy centers in diamond nanocrystals are coupled to whispering gallery modes in a silica microsphere. The composite nanocrystal-microsphere system takes advantage of the exceptional spin properties of nitrogen vacancy centers as well as the ultra high quality factor of silica microspheres. The observation of the normal mode splitting indicates that the dipole optical interaction between the relevant nitrogen vacancy center and whispering gallery mode has reached the strong coupling regime of cavity QED.

Interactions of single atoms with electromagnetic fields in a microresonator have been a central paradigm for the understanding, manipulation, and control of quantum coherence and entanglement [1-4]. Of particular importance is the regime of strong coupling, characterized by a reversible exchange of excitation between an atom and cavity fields. This coherent exchange can induce atom-photon, atom-atom, and photon-photon entanglement and plays a key role in many quantum information processes. Strong-coupling cavity QED has been realized with trapped atoms and with solid-state systems using quantum dots and Cooper pair boxes as artificial atoms [5-9]. A solid-state cavity QED system circumvents the complexity of trapping single atoms in a microresonator and can potentially enable scalable device fabrications.

In order to use the strong-coupling process for quantum control of entanglement in a solid-state environment, the cavity QED system needs to feature a robust spin coherence, since coupling to the surrounding environment leads to rapid decay of most other forms of quantum coherences. Among various solid-state spin systems, nitrogen vacancy (NV) centers in diamond have emerged as one of the most promising candidates. A NV center consists of a substitutional nitrogen atom and an adjacent vacancy in diamond. These defect centers can feature near unity quantum efficiency, a homogenous linewidth as small as 50 MHz, and an electron spin decoherence time exceeding 50 μs at room temperature [10-12]. The remarkable optical and spin properties of NV centers have enabled the realization of single-photon generation as well as coherent Rabi oscillation of single electron spins [13-15]. NV centers formed in diamond nanocrystals retain similar properties as those in bulk diamonds [14, 15].

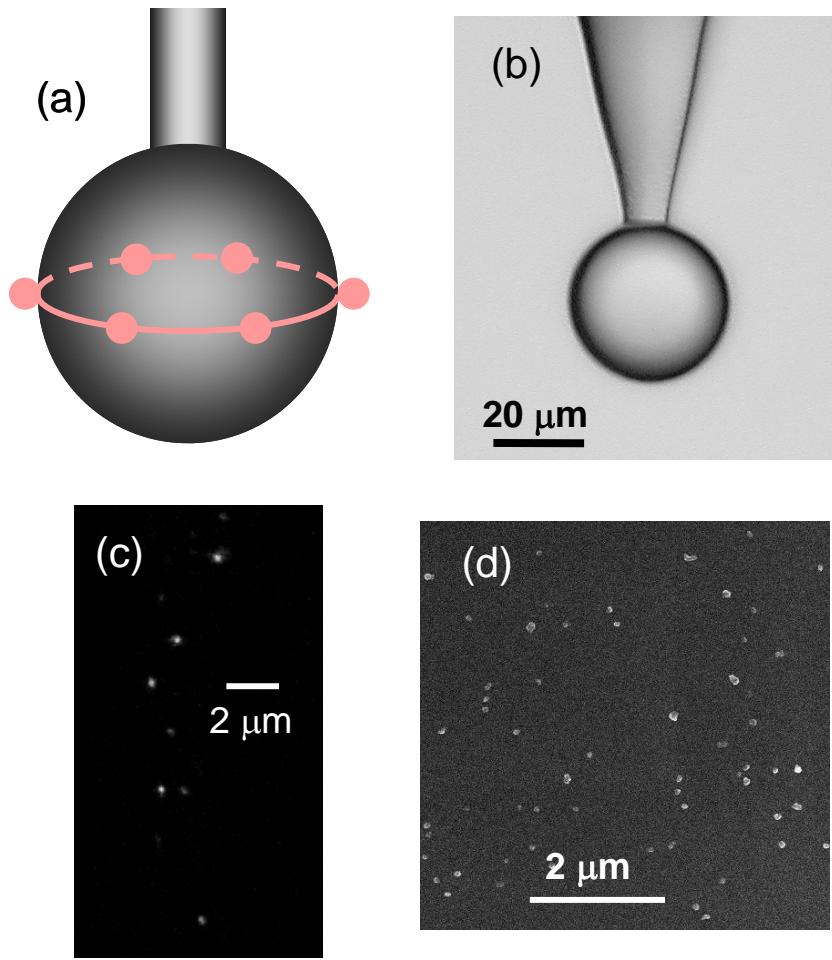

**Figure 1.** (a) Schematic of diamond nanocrystals coupling to a WGM in the equator of a microsphere. (b) An optical image of a deformed silica microsphere system used for the cavity QED measurements. (c) A confocal microscopy image of photoluminescence of NV centers from individual nanocrystals deposited on a silica microsphere. Note that only a ring-shaped section of the sphere surface is in focus. A portion of the ring is displayed in the figure. (d) A scanning electron micrograph image showing diamond nanocrystals deposited near the equator of a silica microsphere.

Here we report the experimental realization of a cavity QED system, in which nitrogen-vacancy (NV) centers in diamond nanocrystals are coupled to a whispering gallery mode (WGM) in a silica microsphere. The composite nanocrystal-microsphere system combines and takes advantage of both the exceptional spin properties of NV centers and the ultra high quality factor of silica microspheres. The normal mode splitting observed in the composite nanocrystal-microsphere system provides experimental evidence for strong-coupling between a WGM and single NV centers in diamond nanocrystals.

Figure 1a shows a schematic of nanocrystals coupling to a WGM at the equator of a microsphere. WGMs in both spherical and toroidal silica microresonators can feature ultra high quality-factors ($>10^8$) along with a small mode volume [16, 17]. These microresonators are deemed to offer the best figure of merit for strong-coupling cavity QED [18-21]. Note that recent experimental studies have also shown evidence of strong coupling between semiconductor nanocrystals and WGMs in lower-Q polymer microspheres [22].

An experimental difficulty for cavity QED with silica microspheres is the excitation of high-Q WGMs at low temperature. While elegant techniques such as tapered optical fibers have been developed [23], these techniques are difficult to implement at low temperature, especially at $\lambda \sim 600$ nm. For low temperature measurements, we have developed a technique of free-space evanescent coupling using nearly spherical but non-axisymmetric microspheres [24]. Critical to our experimental efforts is that this technique allowed us to explore a large number of samples to determine the suitable parameters and conditions for achieving strong-coupling.

As shown in earlier studies, when light rays circulate around the equator in a non-axisymmetric silica microsphere, the angle of incidence becomes closest to the critical angle in areas $45^{\circ}$ away from either the long or short axis, leading to a significant increase in the evanescent tunneling rate as well as the evanescent decay length [24, 25]. WGMs can be excited by focusing a laser beam to these areas with the focal point just outside the sphere surface (see the inset in Fig. 2a). While the enhanced tunneling process degrades the Q-factor of the WGMs, Q-factors approaching $10^8$ can still be obtained if the deformation is kept below 2%.

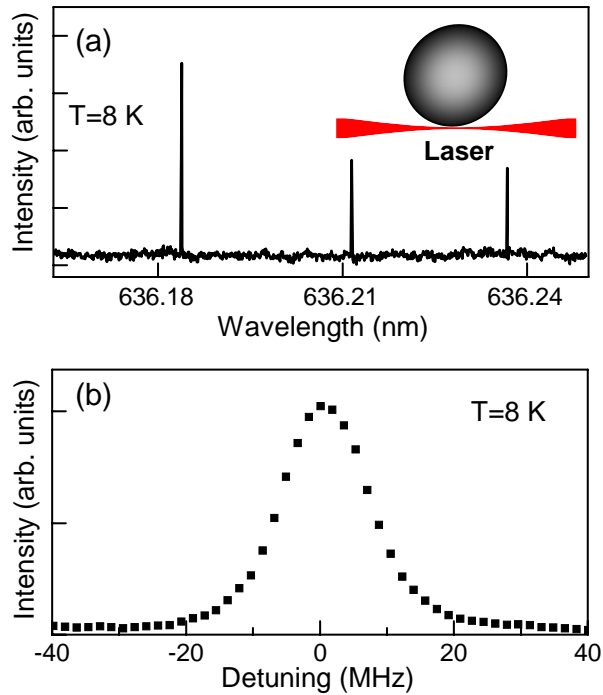

**Figure 2.** (a) WGM spectrum of a nearly spherical but non-axisymmetric silica microsphere with a diameter of 35 µm. The WGMs were excited near the equator with the free-space evanescent tunneling technique shown schematically in the inset (the deformation is exaggerated for clarity of illustration). The mode spacing is 20 GHz. (b) The expanded scan of a WGM resonance shows a cavity linewidth of 14 MHz.

Given the many uncertainties in our attempts to realize strong coupling with NV centers, we have chosen a system that is designed for reaching the strong-coupling regime but is not optimized for achieving the best figure of merit. Specifically, we have used silica microspheres with diameters near 35 μm such that the free spectral range of the WGMs is comparable to the inhomogeneous linewidth of the NV centers (~ 2.5 nm). This allows us to couple WGMs to NV centers both near and away from the center of the inhomogeneous distribution.

Silica microspheres were fabricated by heating a fiber tip with a $CO_2$ laser beam. Non-axisymmetric microspheres were fabricated by fusing together two microspheres with similar sizes. The deformation was gradually reduced through repeated heating to less than 2% and until Q-factors exceeding $10^7$ were observed. Figure 1b shows an optical image of a deformed microsphere used in the cavity QED measurements. For low temperature studies, microspheres were cooled by helium gas in an optical cryostat. A tunable dye laser (linewidth < 1 MHz) frequency-stabilized to an external reference cavity was used to excite WGMs. The power of the excitation laser beam was kept below 0.1 μW. The efficiency of coupling into the WGMs is on the order of 1 %.

With free-space evanescent coupling, we can selectively excite WGMs near the equator of the sphere with principle mode number $p=1$ and with $|m|\sim l$. Approximately 5 to 10 WGM modes, corresponding to modes with slightly different polar angles, can be observed within a given free spectral range. The mode spacing depends on the size and deformation of the sphere and ranges between 10 to 40 GHz. Figure 2a shows the resonant scattering spectrum of a nearly spherical but non-axisymmetric silica microsphere excited near the equator. For these measurements, emissions from the excited WGMs were collected along a direction with minimum background. The intensity of the emission was measured as a function of the excitation laser wavelength. The resonant scattering spectrum obtained probes directly the transmission spectrum of the resonator. Figure 2b shows the expanded scan of a WGM resonance. The linewidth of the WGMs can vary from 6 to 15 MHz, corresponding to a Q-factor as high as $0.8 \times 10^8$.

Type 1b diamond nanocrystals synthesized under high pressure and high temperature and with an average size of 75 nm from de Beer were used in the cavity QED measurements. The nanocrystals were irradiated with electron beam at a dose of $10^{17}/cm^2$ and then annealed in vacuum at 900 K for two hours to create NV centers. Solution deposition was used to attach nanocrystals to a silica microsphere, in which nanocrystals were first dissolved in high purity chloroform and then deposited on the surface of a silica microsphere. In order to minimize the inherent randomness in the solution deposition process, the nanocrystal concentration and the deposition time were adjusted such that after the deposition, linewidths of WGMs near the optical transition frequency of the NV centers were broadened by approximately 20 to 80 MHz at room temperature. The broadening of the WGMs is primarily due to absorption and scattering from the relatively large number of nanocrystals deposited on the sphere surface since additional measurements show that solution deposition without nanocrystals caused negligible degradation in Q-factors.

We have used both confocal optical microscopy and scanning electron micrograph (SEM) to characterize diamond nanocrystals deposited on the sphere surface. Figure 1c shows a confocal microscopic image of photoluminescence from NV centers in individual nanocrystals dispersed on the surface of a silica microsphere through solution deposition. For these measurements, the NV centers were excited with output from an argon ion laser and the image was taken in the spectral range of the NV emission. For the confocal microscopy, only a ring-shaped section of the sphere surface is in focus. A small portion of the ring is displayed in Fig. 1c. Figure 1d shows a SEM image obtained near the equator of a silica microsphere, which allows us to determine the density of the nanocrystals. For the experimental result presented in Fig. 3, the density of nanocrystals near the equator is estimated to be on the order of 1 /$\mu m^2$. Note that a detailed comparison between SEM and confocal microscopy of nanocrystals deposited on a glass slide indicates that a majority of the nanocrystals contains NV centers.

We match a WGM resonance with the optical transition frequency in a NV center by varying the detuning between the WGM and the NV center via temperature tuning. Figure 3 shows the resonant scattering spectra of a composite nanocrystal-microsphere system obtained near $\lambda=634.2$ nm and at temperatures ranging from 6 K to 12.4 K. Note that both the WGM resonance frequency and the NV center transition frequency increase with increasing temperature.

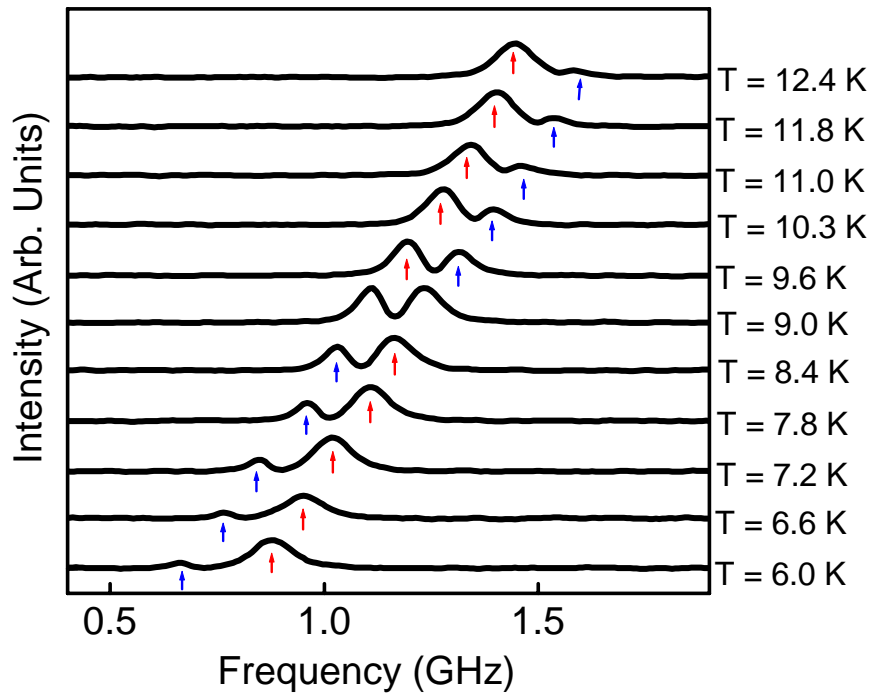

**Figure 3.** Temperature dependence of the resonant scattering spectra of a composite nanocrystal-microsphere system, where the cavity-like and atom-like modes are indicated by red and blue arrows, respectively. The measurements were carried out near $\lambda$=634.2 nm.

Optical excitations of a composite nanocrystal-microsphere system are characterized by the normal modes of the coupled system. For transmission measurements, the cavity-like normal mode dominates the transmission spectrum, while the atom-like normal mode can be observed only when the atom is nearly resonant with the cavity mode, as we will discuss in more detail later. We have used these characteristic behaviors for the assignment of atom-like and cavity-like normal mode resonances. The red arrows in Fig. 3 indicate the dominant cavity-like mode, while the blue arrows indicate the atom-like modes. As shown in Fig. 3, the relative amplitude of the atom-like resonance decreases rapidly as the WGM is detuned from the NV center. Note that unlike cavity QED systems based on the use of semiconductor nanocrystals, the normal mode resonances in Fig. 3 show no noticeable spectral fluctuations under repeated measurements.

For a more detailed analysis of the normal mode coupling in Fig. 3, we plot in Figs. 4a and 4b the expanded scan obtained at T=6 K and T=9 K, for which the WGM and the NV center are detuned and nearly resonant, respectively. Figures 4c and 4d plot the spectral position and linewidth of the two normal modes obtained from Fig. 3 as a function of the temperature. Figure 4c shows the avoided crossing of the spectral positions of the two normal modes. Figure 4d shows that the linewidth of the normal modes approaches the average of the individual linewidth for the NV center and the WGM, when the NV center is nearly resonant with the WGM. Furthermore, a switching of the linewidth between the higher and lower frequency normal modes occurs across the spectral region of the avoided crossing, as expected for normal mode coupling. Note that data obtained for T > 11 K were not included in Fig. 4. For silica microsphere, the WGM resonance frequency becomes nearly independent of the temperature near T=20 K [26]. The temperature dependence of the WGM frequency is nonlinear and levels off as the temperature is increased from 10 K toward 20 K.

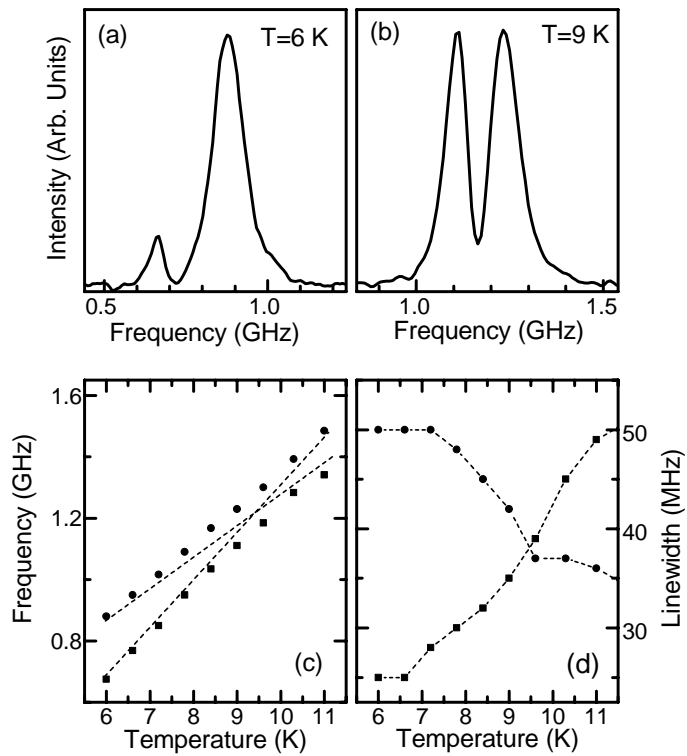

**Figure 4.** (a, b) The expanded scan of normal resonances in Fig. 3 obtained at T=6 K and T=9 K. (c, d) The spectral position and linewidth of the two normal modes as a function of the temperature. The circles and squares are for the upper and lower frequency normal modes in Fig. 3, respectively. The dashed lines are a guide to the eye.

The normal mode splitting when the two normal mode resonances feature nearly equal amplitudes (see Fig. 4b) corresponds to a vacuum Rabi frequency of $g/2\pi = 55$ MHz. Using a radiative lifetime of 12 ns, an effective mode volume of 250 $\mu m^3$, and a Huang-Rhys factor of 3.2 for the zero-phonon resonance of the NV center, we estimate theoretically $g/2\pi \sim 50$ MHz for our cavity QED system, in general agreement with the experimental observation. Figure 5 plots the theoretical linear transmission spectra of a composite atom-cavity system as the detuning between the cavity mode and the two-level atom is varied, where we have used $g/2\pi = 55$ MHz, $\gamma/2\pi = 25$ MHz, and $\kappa/2\pi = 50$ MHz [1]. In particular, the theoretical calculation shows that the relative amplitude of the atom-like mode decreases rapidly as the detuning between the atom and the cavity approaches $\kappa$, in good agreement with the experimental observation. Overall, the normal mode coupling shown in the relative amplitude, spectral position, and spectral linewidth of the two normal mode resonances in Fig. 3 and Fig. 4 demonstrate that the cavity QED system has reached the strong coupling regime.

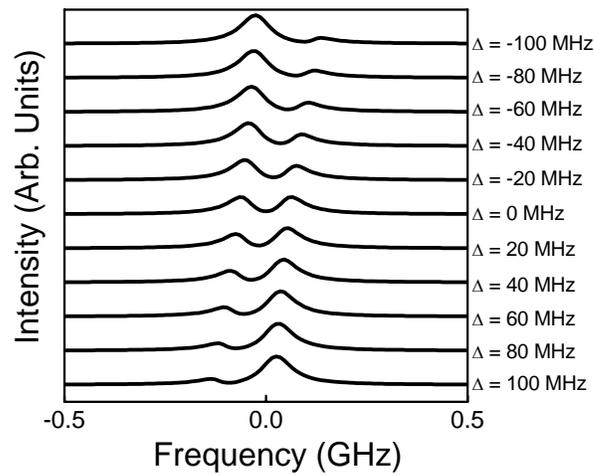

**Figure 5.** Theoretical linear transmission spectra of a composite atom-cavity system, with $g/2\pi$ = 55 MHz, $\gamma/2\pi$ = 25 MHz, and $\kappa/2\pi$ =50 MHz. The detuning between the cavity mode and the two-level atom is indicated in the figure.

It should be added that for composite nanocrystal-microsphere systems with nanocrystal densities much greater than 1 /$\mu m^2$, some WGMs within the inhomogeneous distribution of NV centers feature linewidth as broad as 400 MHz but show no indications of normal mode splitting, while linewidths of other WGMs still remain below 100 MHz (not shown). In these cases, a WGM can couple to multiple NV centers with frequencies clustered within a few hundred MHz. Normal model splitting can only occur when transition frequencies of relevant NV centers are spectrally well separated. While one cannot rule out completely the possibility that the normal mode splitting shown in Fig. 3 may arise from multiple NV centers, statistically, it is highly unlikely that multiple NV centers have a nearly identical transition frequency and, at the same time, transition frequencies of all other NV centers are more than a few hundred MHz away from this frequency.

We have carried out experimental studies on a large number of composite nanocrystal-microsphere systems. On average, when the density of the nanocrystals deposited on the sphere surface is on the order of 1 /$\mu m^2$ and the cavity linewidth is narrower than 100 MHz, we can observe normal mode splitting in approximately one out of four microsphere systems. For an atom-cavity detuning range of 0.5 GHz and an average mode spacing between neighboring WGMs of 20 GHz, the probability that a single NV center can be tuned to be resonant with a WGM near the equator is approximately 0.05. With an effective surface area of 200 $\mu m^2$ for the relevant WGMs, there are on average 200 nanocrystals that can in principle couple to the WGMs near the equator and there are typically multiple NV centers in a given nanocrystal [14]. The relatively small success rate for observing normal mode splitting thus indicates that only a very small percentage of nanocrystals contain spectrally stable NV centers that are suitable for achieving strong coupling.

We note that while the observation of normal mode splitting provides good evidence for the strong coupling between a WGM and single NV centers, experimental studies such as vacuum Rabi oscillation in the photon correlation can provide additional confirmation and valuable information on the underlying strong coupling process. Earlier reports of strong coupling in solid state systems have not included the photon correlation studies. For the composite nanocrystal-microsphere system, the photon correlation studies can be carried out on optical emissions from the phonon sidebands of the NV centers. A major modification of our current experimental setup, however, is needed in order to obtain adequate photon counts for the measurement of the phonon sidebands. In addition, the parameters of our cavity QED system can also be further improved. For example, the cavity decay rate can be greatly reduced by selecting and implanting single nanocrystals with selected transition frequencies into a silica microresonator with nanomanipulation techniques [27]. Toroidal silica resonators can also be used to further reduce the cavity mode volume and thus increase the vacuum Rabi frequency [20].

The realization of a strong-coupling cavity QED system with NV centers and silica microspheres opens the door to employing robust electron spin coherence in the strong-coupling cavity QED regime. For example, an additional microwave pulse can be used to manipulate spin states in the ground-state triplet of a NV center. Strong-coupling cavity QED systems with NV centers featuring µs spin decoherence time at room temperature provide us a highly promising system to realize important quantum information processes such as distributed quantum networks and spin entanglement [28-30].

We thank S.C. Rand for valuable discussions and for providing us NV diamond crystals. This work has been supported by ARO and by NSF.

Table of contents image:

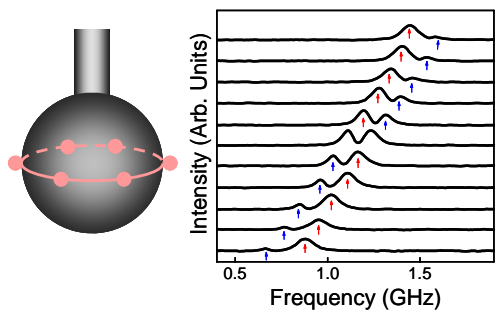